\documentclass[aps,prl,amsmath,amssymb,twocolumn,nofootinbib, showpacs, showkeys, nofootinbib]{revtex4}
\usepackage{pslatex,graphicx,dcolumn,bm,natbib}
\usepackage{multirow}
\usepackage{subfigure}
\date{\today}

\begin{document}

\title{Fluctuations of particle motion in granular avalanches \\
-- from the microscopic to the macroscopic scales}

%% Notice placement of commas and superscripts and use of &
%% in the author list

\author{Ziwei Wang$^1$, and Jie Zhang$^{2,3,\ast}$}

\email{jiezhang2012@sjtu.edu.cn}
\affiliation{$^1$Zhiyuan College \& $^2$Institute of Natural Sciences \& $^3$Department of Physics and Astronomy, Shanghai Jiao Tong University, Shanghai 200240, China
}

\begin{abstract}
 In this study, we have investigated the fluctuations of particle motion, i.e. the non-affine motion, during the avalanche process, discovering a rich dynamics from the microscopic to the macroscopic scales. We find that there is strong correlation between the magnitude of the velocity fluctuation and the velocity magnitude in the spatial and temporal domains. The possible connection between this finding and STZ is discussed based on the direct measurement of the T1 events. In addition, the velocity magnitude of the system and the stress fluctuations of the system are strongly correlated temporally. Our finding will pose challenges to the development of more rigorous theories to describe the avalanche dynamics based on the microscopic approach. Moreover, our finding presents a plausible mechanism of the particle entrainment in a simple system.
\end{abstract}

\keywords{Granular avalanche, non-affine motion, T1 event, stress fluctuation}

-\pacs{83.80.Fg, 45.70.-n, 81.05.Rm, 61.43.-j}

\maketitle

Granular avalanches are ubiquitous in nature. A simple demonstration will be a continuous tilting of a pile of granular materials, eventually causing an avalanche with a massive rearrangement of particles to self-organize the system to a new stable configuration at a lower energy state \cite{jaeger1996granular, daerr1999two}. The study of such processes has important applications in geophysics, in mechanical and civil engineering, in pharmaceutical applications, and in studying the natural disasters such as the landslide and mud-slide, et al \cite{ottino2000mixing,christen2010ramms,hungr2004entrainment,iverson2011positive}. Historically, the study of the avalanche of sand pile has played an important role in the development of important theoretical concepts such as the Coulomb's laws of friction \cite{duran2000sands} and the self-organized criticality \cite{bak1987self,jaeger1989relaxation,frette1996avalanche}, et al.

In recent years, granular avalanches have become a field of intense research activities. These include: (1) the studies of the critical behavior at the onset of an avalanche \cite{sen1994onset, bretz1992imaging, ramos2009avalanche, Dahmen_NatPhys}; (2) the numerical analysis of the  bimodal contact network contributions to the instability of the avalanche \cite{staron2005friction}; (3) the continuum theory description of the surface flow profile of the avalanche \cite{bouchaud1995hysteresis, boutreux1998surface, aradian1999thick} and the extended theory \cite{makse1999continuous} and the combined theoretical/experimental analysis for particle segregation and for surface wave instability analysis \cite{koeppe1998phase, aranson2001continuum, aranson2006transverse, Bonneau2007}; (4) the novel dynamics caused by particle shapes \cite{borzsonyi2005two}, particle cohesion \cite{tegzes2002avalanche}, and the variation of volume fractions \cite{gravish2014effect}; (5) the bifurcation of flow dynamics and instabilities due to the change of the external driving \cite{amon2013granular, fischer2009transition, zimber2013polydirectional}; (6) the the self-similarity of the surface mean velocity profile \cite{du2005instantaneous} and correlations between the starting and the ending angles \cite{fischer2008dynamics}; (7) the slow relaxation dynamics following an avalanche \cite{Deboeuf2003jamming} and the diffusive particle motion in the steady avalanche \cite{KHill_PRL}. Nevertheless, there is still a lack of a coherent physical picture which connects the macroscopic/mesoscopic dynamics of a granular system to the microscopic dynamics at the particle scale in the granular avalanche process \cite{aranson2006patterns}.

In this study, we try to fill the missing link between the microscopic and the macroscopic/mesoscopic dynamics by focusing much of the attention on analyzing the non-affine particle motion from the particle scale to the system scale during granular avalanches in a quasi two dimensional (2D) system -- a rotating drum partially filled with a 2D layer of granular disks to create avalanches, as sketched in Fig. 1 (See Methods for details). Here we show that the magnitude of velocity $v$ and the non-affine velocity $\delta v$ are strongly correlated in the temporal domain from the microscopic particle scale to the macroscopic system scale. In addition, we also find strong correlations of these two fields in the spatial domain. This is surprising since normally one would expect the fluctuations of particle motion become weaker as the scale increases. The strong spatial correlation between $\delta v$ and $v$ can be partially understood from the numerical model of a stochastic steady shear flow. Moreover, we have successfully observed T1 events, the cores of STZ in 2D systems, for the first time in a granular experiment. We will discuss the possible connection between STZ and the above strong correlations of the two fields $\delta v$ and $v$. The fluctuations of particle velocity can have profound influence on the macroscopic stress changes of the system.  We have discovered a connection between the velocity dynamics of particles and the fluctuation of the total stress of the system from the perspective of momentum transfer. Our finding may pose a challenge to the further development of the theoretical description of avalanche dynamics based on microscopic dynamics of particles. Our finding will be important to some geological processes such as particle entrainment and bed erosion in snow avalanches and landslides.

This paper is organized as follows. In the Results Section, we present experimental results along with the discussion, which include the spatial and temporal correlations of velocity and velocity fluctuations, the numerical simulation, STZ and T1 events, and the stress fluctuations of the system. In the Discussion Section, we will draw conclusions.

\section{Results}

\subsection{A. Velocity fluctuations}
\begin{figure}
\centerline{\includegraphics[width=0.5\textwidth]{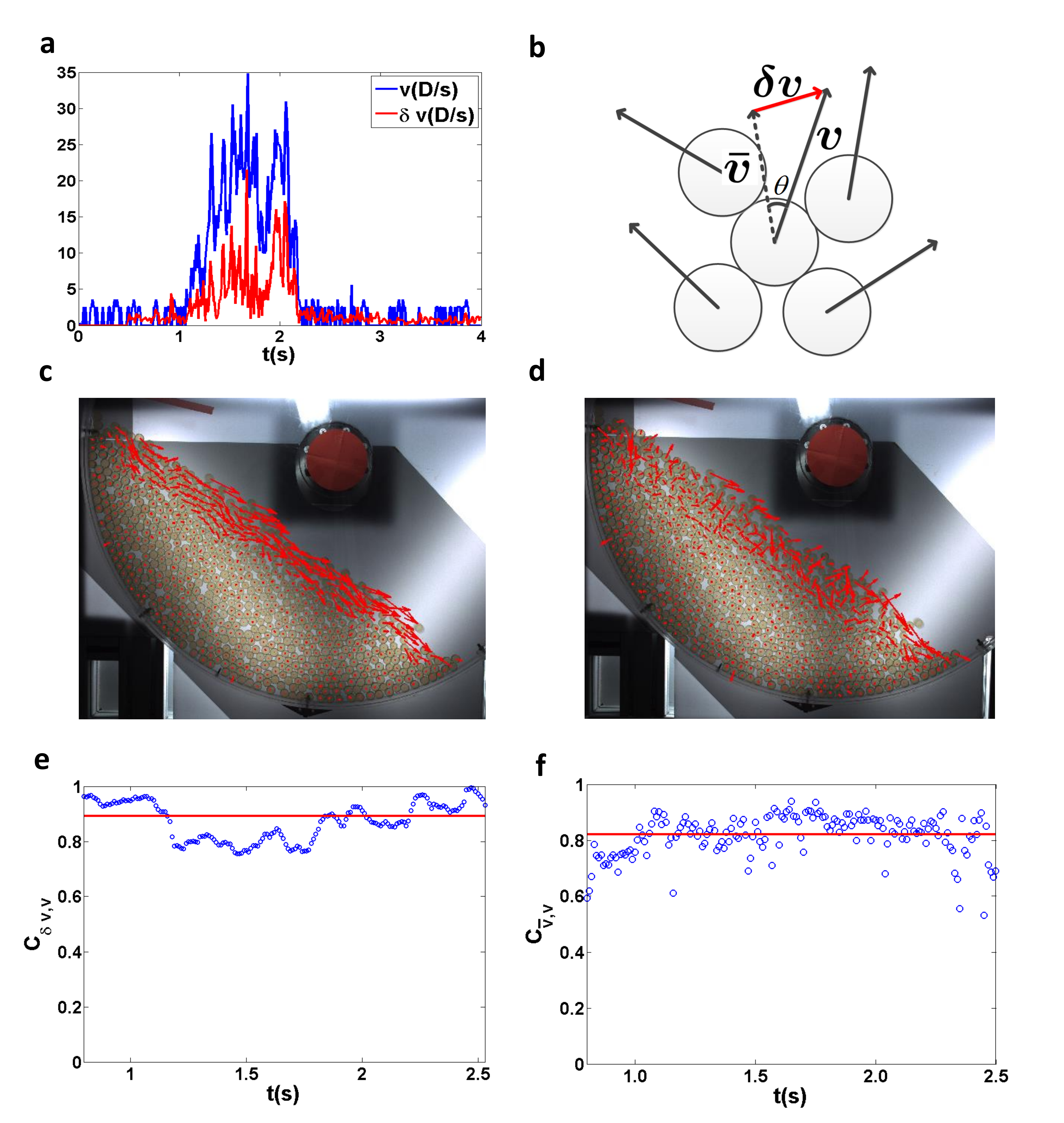}}
\caption{\label{fig:figure2} (a)The evolution of velocity $v$ and its fluctuation $\delta v$ of a particle in one avalanche. (b) A schematic of the definition of velocity fluctuations $\delta {\bf v}$ and the deflection angle $\theta$ between the local mean velocity $\bar{{\bf v}}$ and ${\bf v}$. (c) Vector field ${\bf v}$ at $t=1.4s$ in one granular avalanche. (d) The corresponding vector field of $\delta {\bf v}$. In (c-d), vectors are drawn using red arrows on the same size scale. (e) The correlation $C_{\delta v, v}$ as a function of time during the avalanche. (f) The correlation $C_{\bar{v},v}$.}
\end{figure}

We first analyze the evolution of the local and the global velocity fluctuations. In Fig.~\ref{fig:figure2}(a), we plot $\delta v$ (red) and $v$ (blue) versus time respectively for an arbitrary particle, which shows that the curves of $\delta v$ and $v$ have a similar trend and thus may be correlated. Note that the uncertainty of the velocity measurement is $2.5D/s$ with $D$ being the average diameter of a particle. The average temporal correlation between $\delta v$ and $v$ of individual particles is around $0.79$ of all runs, which is reasonably strong. Here the velocity fluctuation of a given particle $\delta {\bf v}$ = ${\bf v}-\bar{{\bf v}}$, as illustrated in Fig.~\ref{fig:figure2}(b). Note that in this paper we use bold fonts for vectors, e.g. $\delta {\bf v}$ and ${\bf v}$, and regular fonts for moduli, e.g. $\delta v$ and $v$. We define $\theta$ as the deviation angle between $\bar{\bf v}$ and ${\bf v}$. The local mean velocity $\bar{\bf v}$ is measured by averaging local velocities of particles centered around the given particle using a Gaussian coarse-graining function with a standard deviation of $2D$  \cite{goldenberg2006continuum,zhang2010coarse}. A sample image of $\bf v$ and the corresponding sample image of $\delta {\bf v}$ are shown in Fig.~\ref{fig:figure2}(c-d). The dynamics of the spatial correlations between $\delta v$ and $v$ are analyzed as shown in Fig.~\ref{fig:figure2}(e), where the correlation function $C_{\delta v, v}$ reveals a strong spatial correlation of the two quantities during the whole avalanche process. There are random fluctuations around the average, close to 0.89, as indicated using the red horizontal line in this figure. A strong correlation $C_{\bar v, v}$ between the local mean velocity $\bar v$ and $v$ is also shown in Fig.~\ref{fig:figure2}(f). Table~\ref{table:table1} summarizes the statistics of the average and the standard deviation of $C_{\delta v, v}$ and $C_{\bar v, v}$ from ten different runs, which are consistent for all runs.
%table 1
\begin{table}[h]%\footnotesize
{
\caption{\label{table:table1}
The statistics of the spatial correlations $C_{\delta v,v}$ and $C_{\bar{v},v}$ in ten different runs. Here $\delta v$ is the magnitude of the nonaffine velocity of a particle, $v$ is the magnitude of the particle velocity and $\bar{v}$ is the magnitude of the local mean velocity at the center of the particle.
}
\scriptsize
\begin{tabular}{|c|c|c|c|c|c|c|c|c|c|c|c|c|}
\hline
\multicolumn{12}{|c|}  {avalanche number $n_0$} \\
\hline
  & $n_0$ & 1 & 2 & 3 & 4 & 5 & 6 & 7 & 8 & 9 & 10  \\ \hline
\multirow{2}{*} {$C_{\delta v, v}$} & mean &  0.73  & 0.89  & 0.87 & 0.90 & 0.84 & 0.84 & 0.63 & 0.92 & 0.70 & 0.84	\\ 	
               & std & 0.14 & 0.047 & 0.14 & 0.054 & 0.10 & 0.20 & 0.083 & 0.057 & 0.10 & 0.10	\\
 												   \hline
\multirow{2}{*} {$C_{\bar{v}, v}$} & mean &  0.85  & 0.91  & 0.89 & 0.92 & 0.77 & 0.74 & 0.86 & 0.67 & 0.88 & 0.80	\\ 	
               & std & 0.070 & 0.034 & 0.067 & 0.058 & 0.095 & 0.10 & 0.046 & 0.090 & 0.042 & 0.072	\\
 												   \hline
\end{tabular}

}
\end{table}

\begin{figure}
\centerline{\includegraphics[width=0.5\textwidth]{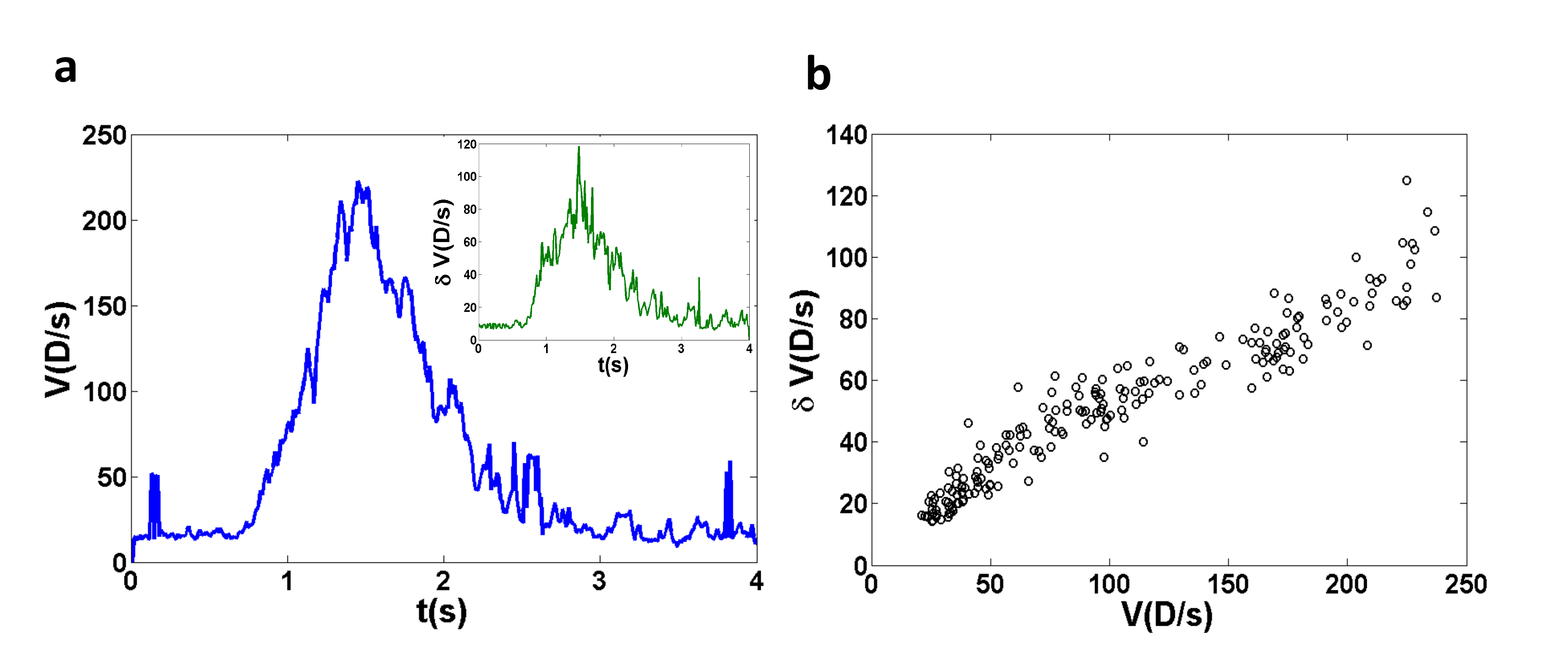}}
\caption{\label{fig:figure3} (a) The global velocity $V$ (main plain) and global non-affine velocity $\delta V$ (inset) versus time. Here $V=\sqrt{\sum_i v_i^2}$ and $\delta V=\sqrt{\sum_i \delta v_i^2}$. (b) A scatter plot of $\delta V$ versus $V$ produced from the results of (a).}
\end{figure}

We define the global velocity fluctuation $\delta V$ and global velocity $V$ as $\delta V=\sqrt{\sum_i \delta v_i^2}, V=\sqrt{\sum_i v_i^2}$, and plot them in Fig.~\ref{fig:figure3}(a). Clearly, their time evolution is similar and the scatter plot of the two variables is shown in (b). The strong correlation between $\delta V$ and $V$ is unexpected since normally the velocity fluctuations will become much weaker at the system scale. This means that the ordered directional movement of all particles can hardly occur in granular avalanches. Note that the above observations are quite different from the thermodynamical systems under phase transitions, e.g. melting of a crystal where fluctuations increase as the system becomes more disordered without correlated macroscopic motion \cite{landau1980statistical, kadanoff2000statics}. In continuum theory description of granular flows of avalanche, fluctuations are usually ignored \cite{bouchaud1995hysteresis, boutreux1998surface, aradian1999thick, aranson2006patterns}. However, our findings suggest that ignoring such fluctuations might be problematic especially for applications such as particle entrainment or bed erosion \cite{christen2010ramms,hungr2004entrainment,iverson2011positive} where mesoscopic particle dynamics is critical. This may pose a challenge for the further development of theoretical descriptions of the dynamics of granular avalanches\cite{goldenberg2006continuum}.

\begin{figure}
\centerline{\includegraphics[width=0.5\textwidth]{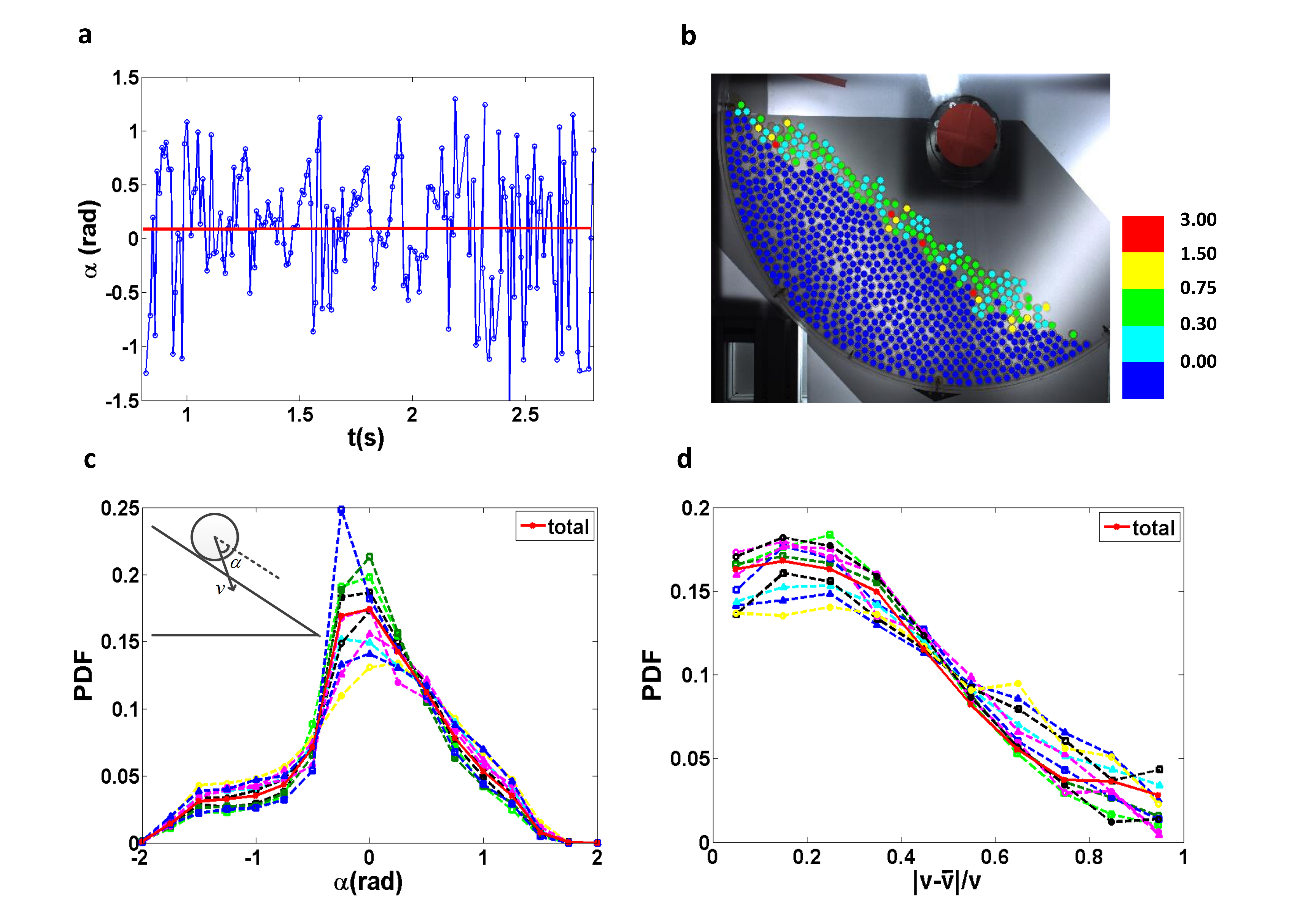}}
\caption{\label{fig:figure4} (a) $\alpha$(t) of a moving particle during one avalanche. Red line is the average of $\alpha$(t), which equals 0.10. (b) The spatial distribution of $\alpha$ at $t=1.5s$ in one avalanche. (c) The distributions of $\alpha$ of different runs. Here the solid line is the total distribution of all runs. Here the positive $\alpha$ is clockwise with respect to the inclination of the surface (main plain). A schematic of the definition of $\alpha$, which is the deviation angle of $v$ with respect to the inclination of the surface (inset). (d) The distributions of $\frac{|v-\bar{v}|}{v}$ of all moving particles in different runs. The solid red line is the total distribution of all runs.}
\end{figure}

To better characterize the fluctuations of the moving particles, in addition to $\delta v$, we measured the deviation angle $\alpha$ between ${\bf v}$ and the inclination. Figures~\ref{fig:figure4}(a-b) present the temporal fluctuation of $\alpha$ of a particle and the spatial distribution of $\alpha$, showing no discernible correlation with either $v$ or $\delta v$.  The PDF of $\alpha$ is shown in Fig.~\ref{fig:figure4}(c), which is robust against different runs -- suggesting an intrinsic disordered characteristics of the avalanche. Similarly, the histograms of $(v-\bar{v})/v$ of different runs are similar as shown in Fig.~\ref{fig:figure4} (d), suggesting a possible connection with $\alpha$.

\subsection{B. A numerical model}
To get some understanding of the above results, we have constructed a numerical model where we simplify the particle motion using a 2D steady shear flow with stochastic orientations of local velocity compared to the inclination. This shear flow is constructed by assigning a velocity vector field on a 200$\times$10 square lattice according to {\bf v} $\propto e^{-z/z_0} \hat{x} $ where $z$ represents the vertical coordinate and $\hat{x}$ is the unit vector of the horizontal direction. Here the lattice is used only for the simple presentation. In fact the velocity is treated as a field here and only the position z matters. The stochastic motion of the particle is then introduced by setting the orientation angle $\alpha$ of each particle's velocity with respect to $\hat x$ on every grid point according to a priori exponential distribution of $\alpha$ , $P(\alpha)=0.5\lambda e^{-\lambda |\alpha|}$. Here the parameter $\lambda$ is set to equal 1.5 as estimated from Fig.~\ref{fig:figure4} (c).

\begin{figure}
\centerline{\includegraphics[width=0.5\textwidth]{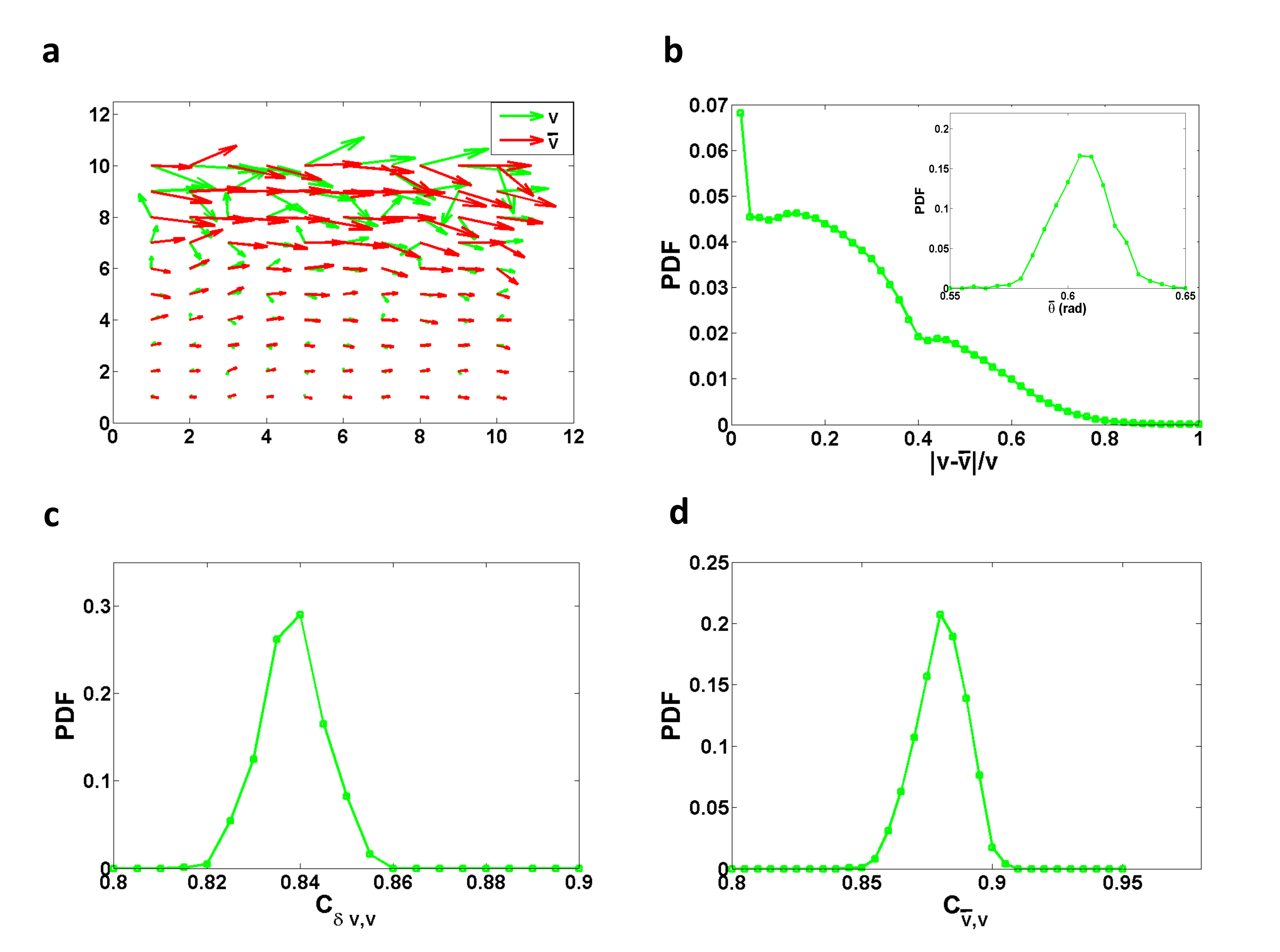}}
\caption{\label{fig:figure5} (a) A portion of an artificially created flow field within a total of 200 $\times$ 10 grid size when $\alpha$ is assumed to obey an exponential distribution, i.e. $P(\alpha)=0.75e^{-1.5|\alpha|}$. Here the flow field is drawn on a square lattice with green arrows representing for the velocity {\bf v} on each grid point and red arrows for the local coarse-grained mean velocity $\bar{{\bf v}}$. (b) Distribution of $\frac{|v-\bar{v}|}{v}$ (main plain) and $\bar{\theta}$ (inset), where $\bar{\theta}$ is the spatial average of all grid points in each simulation. (c) Distributions of $C_{\delta v, v}$, where $C_{\delta v, v}$ is the spatial correlation of $v$ and $\delta v$ of each simulation. (d) Distributions of $C_{\bar{v}, v}$, where $C_{\bar{v}, v}$ is the spatial correlation of $v$ and $\bar{v}$ of each simulation. Here in (b-d) the statistics are from 1000 simulations.}
\end{figure}

A small portion of the velocity field ${\bf v}$ is shown in Fig.~\ref{fig:figure5} (a) in green color generated using $P(\alpha)$. Along with the ${\bf v}$ field, at each grid point the local mean velocity field $\bar{\bf {v}}$ can be computed following the exactly same Coarse-Graining method in analyzing the experimental results. The $\bar{\bf {v}}$ field are drawn in red color in Fig.~\ref{fig:figure5}(a). We then compute the statistics of the relevant physical quantities -- $\frac{v-\bar{v}}{v}$, $\bar{\theta}$, $C_{\delta v, v}$, and $C_{\bar{v},v}$ in order to compare the results with the experiment, as shown in Fig.~\ref{fig:figure5}(b-d). In Fig.~\ref{fig:figure5}(b) the distribution of $\frac{v-\bar{v}}{v}$ is broad, with main features qualitatively similar to Fig.~\ref{fig:figure4}(d) except at small $\frac{v-\bar{v}}{v}$ where it is beyond the meaningful resolution of the experimental measurement . In inset of Fig.~\ref{fig:figure5}(b), $\bar{\theta}$ shows an average value around 0.61, in good agreement with the experiment. In Fig.~\ref{fig:figure5}(c-d), the average of $C_{\delta v,v}$ and $C_{\bar v,v}$ are around $0.84$ and $0.87$ respectively, comparable to the experimentally measured values in Table~\ref{table:table1}.

Figure~\ref{fig:figure5} shows that our numerical model agrees qualitatively and to some extent quantitatively with the experimental measurement. However, this model is oversimplified by ignoring all the dynamics. In fluid mechanics, the correlation between the local mean flow and the fluctuations has been known for quite a long time perhaps since the discovery of the Taylor dispersion effect \cite{GITaylor}. In granular experiments, similar effects have been observed in the steady flows in a Couette experiment \cite{Utter_PRE04} and also in a rotating drum experiment \cite{KHill_PRL}, where diffusive particle motion has been tracked, showing the correlations between local mean shear flow and the diffusion perpendicular to the shear. However, the perspective is different: these existing experiments take a Lagrange point of view to track diffusive particle motion by following particles in a sufficiently long time in a steady shear flow \cite{Utter_PRE04, KHill_PRL}; whereas in our experiment, in a Eulerian reference frame, we have found a much 'stronger' result by decomposing affine and nonaffine motion of particles at a given time in a non-steady shear flow. On the application side, the correlations between $\delta v$ and $v$ can enhance the fluidization of the stable granular configurations under the surface particle flow, which might be important to geological processes, such as particle entrainment and bed erosion during avalanches \cite{christen2010ramms,hungr2004entrainment,iverson2011positive}.

\subsection{C. STZ and T1 events}
The strong correlation between local mean velocity and velocity fluctuations may seem to be a direct consequence of the STZ theory, which is a beautiful theory in description of the plasticity in amorphous solid \cite{LangerReview, LangerPRE}.However, the direct application of the STZ to the present system is highly nontrivial . The reasons are the following. First, STZ is based on the assumption that ¡°the material of interest is solid-like¡± \cite{LangerReview, LangerPRE}, whereas during the avalanche, the surface layer particles flow like a liquid. Second, STZ can naturally lead to the prediction of a material¡¯s yield stress, which may not be the relevant parameter in granular avalanche; the key parameter is a dimensionless ratio, i.e. the angle of repose. Third, STZ typically assumes that there is a separation of time scales, which is not valid in our experiment. Moreover, STZ is based on the assumption that localized rare plastic events control the dynamics of the system, which should be violated in the existence of a system-spanning event such as the avalanche. Lastly, the existence of force-chains may cause long-range, nonlocal interactions, and the athermal nature of the present system will may add additional complications.

\begin{figure}
\centerline{\includegraphics[width=0.5\textwidth]{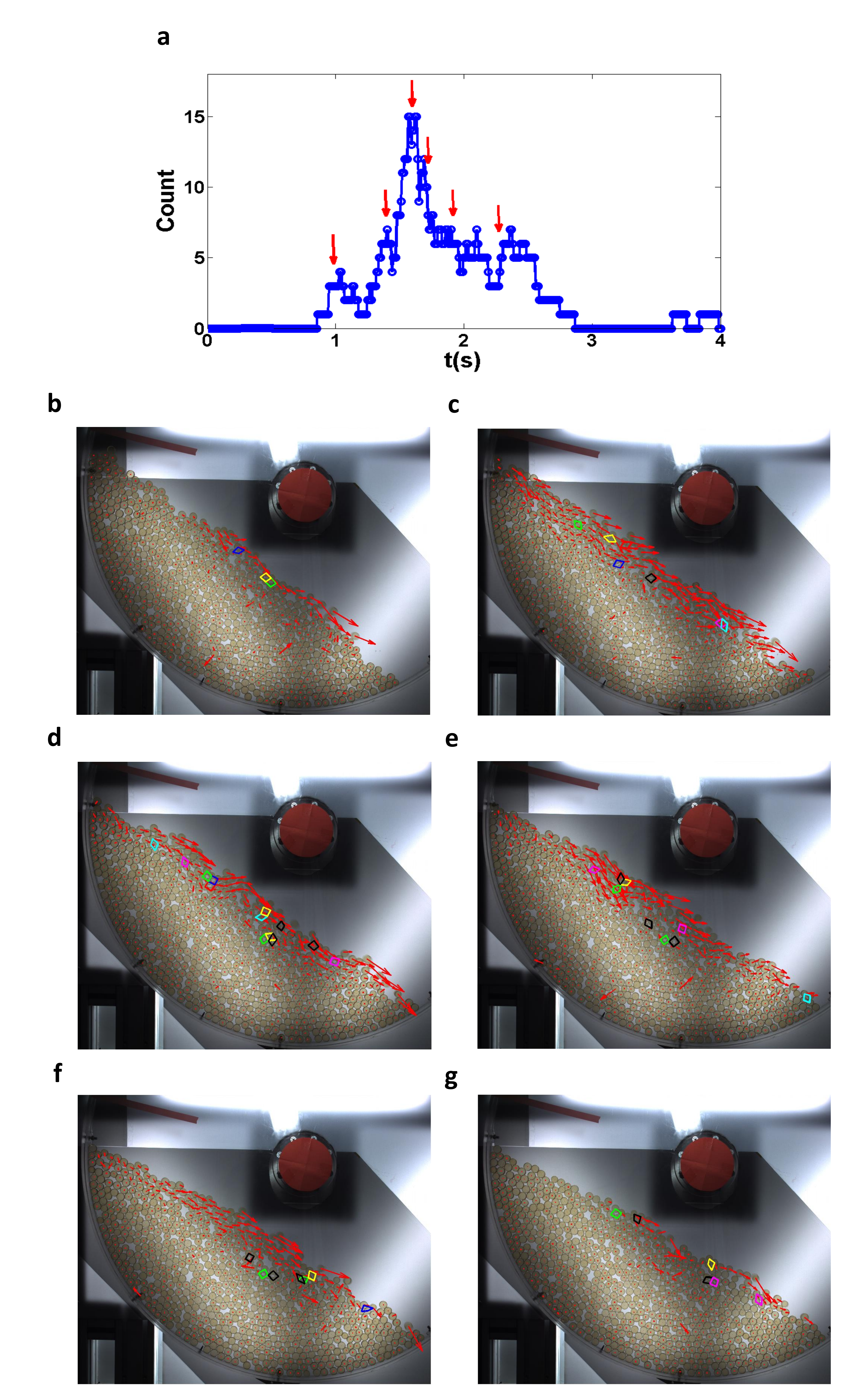}}
\caption{\label{fig:figure6} (a) The population of T1 events versus time during one avalanche. The six red arrows represent the six different times shown in (b-g). (b-g) Snapshots of the T1 events and velocity field during one avalanche at t=1.0 s, 1.4 s, 1.6 s, 1.7 s, 1.9 s, and 2.3 s, respectively.}
\end{figure}

Nonetheless, to our great surprise we do find T1 events, which are the cores of STZ in a 2D system \cite{LangerPRE}. The results are shown in Fig.~\ref{fig:figure6}, where panel (a) plots the evolution of total population of T1 events; comparing to Fig.~\ref{fig:figure3}(a), the trend is similar. The remaining six panels of Fig.~\ref{fig:figure6} show the snapshot of spatial distributions of T1 events with time instants marked with arrows in Fig.~\ref{fig:figure6}(a). The spatial distributions of the T1 events, i.e. STZ, are aligned with the inclination. Compared with Fig.~\ref{fig:figure3}(a), when $V(t)$ of the system increases, T1 events start to locate deeper into the inclination, possibly because that STZ is a description of plastic deformation. But due to the coexistence of flow regime and STZ, it is difficult to evaluate which one plays the dominant role in controlling the dynamics of the system. Most likely, it is the interplay of both flow regime and STZ: an STZ randomly dies and gets reborn somewhere else due to the noise of the flowing particles; on the other hand, STZ causes more surrounding particles to be unstable and flow. Considering the small number of T1 events, the correlations of local mean velocity and velocity fluctuations are more likely an intrinsic property of particle flow. The T1 events, and thus STZ's, have been successfully observed in 2D experiments of foams \cite{DenninPRL97,DenninPRE04, DenninPRE06, DenninPRE08,KablaPRL03,DOlletJFM07,ChenSM12}; to our knowledge there is no direct experimental observation of T1 events in 2D granular experiments in literature. It might be deeply connected with the frictional properties of granular particles, causing frustration of particle motion in the T1-event-like deformation. We speculate that the direct observation of T1 events in the present system is due to the fluidization of the particles at the inclination of the surface, causing the particles at the plastic zones to act like effective frictionless particles. How to extend the STZ theory to such a regime is something interesting and potentially important, but is beyond the scope of the present study.

\subsection{D. Stress fluctuations}

\begin{figure}
\centerline{\includegraphics[width=0.5\textwidth]{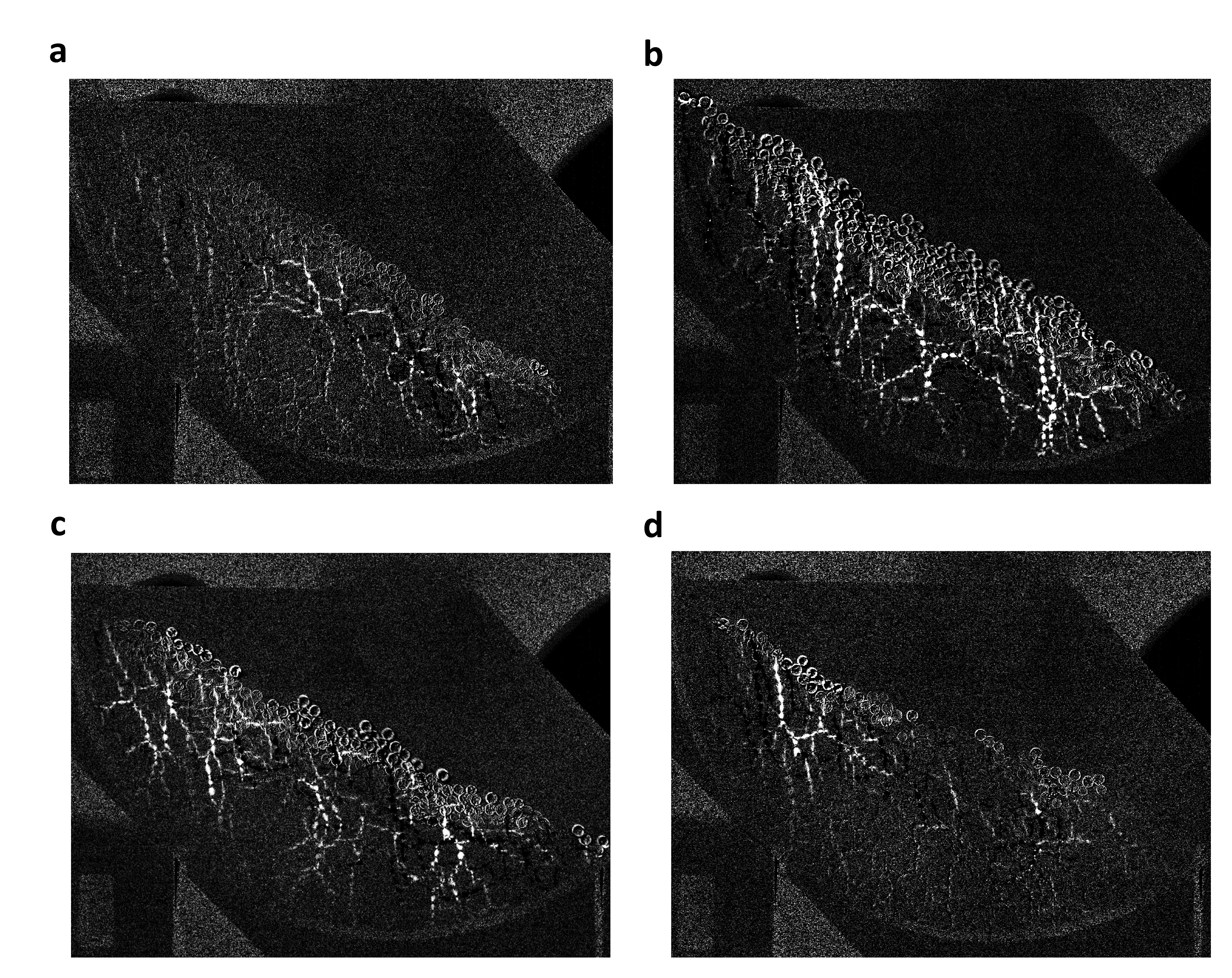}}
\caption{\label{fig:figure7} (a-d) Snapshots of the change of the force network during one avalanche at t=1.0 s, t=1.4 s, t=2.0 s, and t=2.4 s, respectively.}
\end{figure}

\begin{figure}
\centerline{\includegraphics[width=0.5\textwidth]{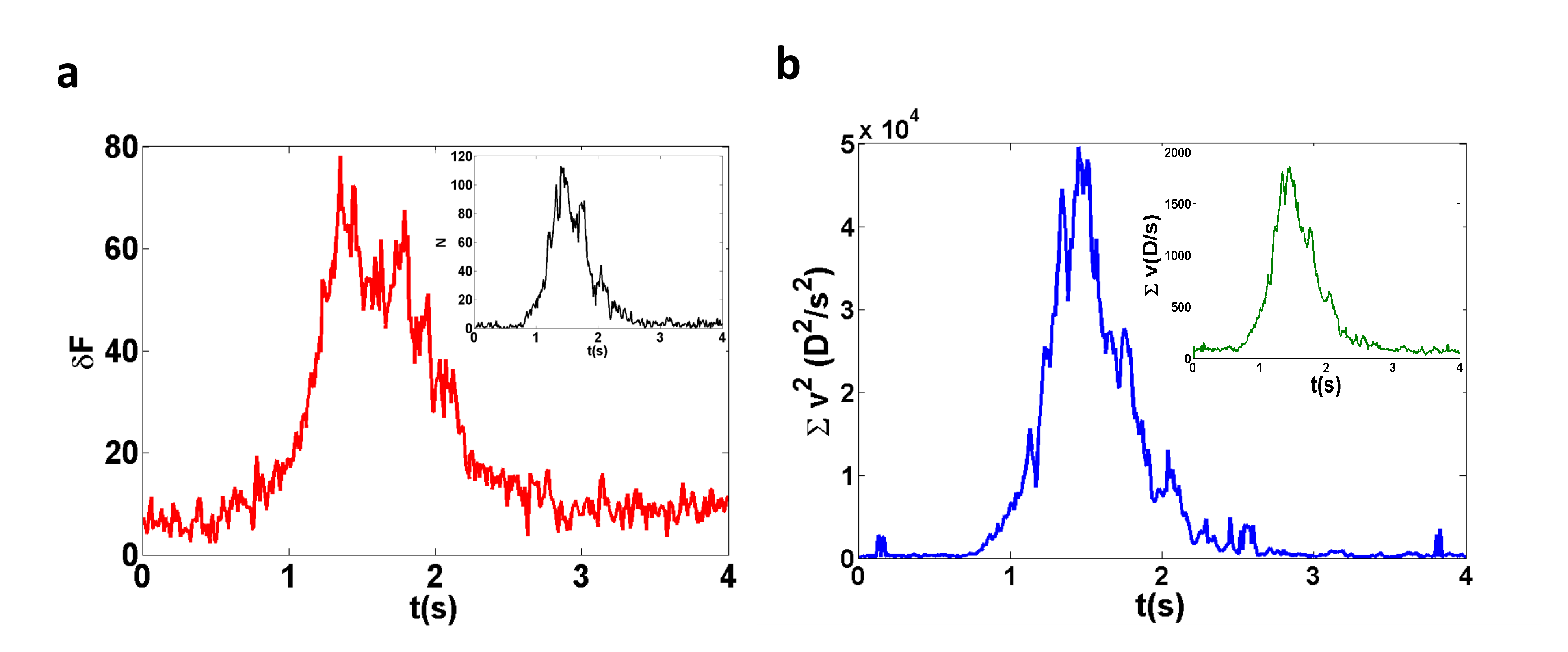}}
\caption{\label{fig:figure8} (a) The magnitude of the change of the stresses of all particles from two neighboring frames of images, $\delta F$, as a function of time during the avalanche. The frame rate is $100 frames/sec$. The inset shows the number of avalanche particles $N$ versus time.(b) The summation of $v^2$ (main plain) and $v$ (inset) versus time during the avalanche.}
\end{figure}

Finally, we try to make a connection between the velocity $v$ of individual particles and the change of stress for the whole system -- a reflection of the energy dissipation and transformation. Here we measure the change of internal stress $\delta F$ within the system by the means of $G^2$ (See e.g. Ref\cite{clark2014collisional}), which computes the square of the gradient of the differential image between two stress images in successive time frames, as shown in Fig.~\ref{fig:figure7}. $\delta F$  versus time is plotted in Fig.~\ref{fig:figure8} (a). In Ref\cite{clark2014collisional}, a simple model is proposed to describe an impact process where the intruder transfers momentum to the granular materials through a sequence of random collisions with the force network. For our system, during an avalanche there are random collisions between the flowing particles at the surface layers and massive stable particle configurations at the bottom where the kinetic energy is dissipated and the momentum is transferred to excite new force-chain networks. We first analyse the collision between a single particle and the bottom as shown in the inset of Fig.~\ref{fig:figure4} (c). The change of $\delta F$ is connected with the momentum transfer perpendicular to the surface layers by the collision, which imparts momentum $\Delta p=(1+e)\frac{mM}{m+M}v sin\alpha$, where $\alpha$ is the angle between $v$ and the inclination of the surface, $e$ is the restitution coefficient, and $m$ and $M$ are the mass of a single particle and the bottom configuration, respectively. The typical collision time can be approximated as $\Delta t=\frac{\gamma D}{v sin\alpha}$, where $D$ is the diameter of a particle and $\gamma$ is a constant. Thus, the average force acting on the bottom perpendicular to the surface is
\begin{equation}
f=\frac{\Delta p}{\Delta t}=\frac{(1+e)v^2 sin^2\alpha}{\gamma D}(\frac{mM}{m+M})
\end{equation}
To obtain the total stress change of the system, we need to sum the forces over all moving particles, i.e.
\begin{equation}
\delta F=\sum_i f_i=\frac{(1+e)mM}{\gamma D(m+M)}\sum_i v_i^2 sin^2\alpha_i
\end{equation}
The above expression can be simplified and lead to
\begin{equation}
\langle \delta F \rangle=C\sum_i \langle v_i^2 \rangle \langle sin^2\alpha_i \rangle =C\langle sin^2\alpha\rangle \sum_i \langle v_i^2\rangle
\end{equation}
if $\alpha$ and $v$ are completely uncorrelated, where $C=\frac{(1+e)mM}{\gamma D(m+M)}$ and $\langle sin^2\alpha\rangle$ are constant. This is a nontrivial requirement. As shown on the histograms of $\alpha$ of all runs in Fig.~\ref{fig:figure4} (c), $\alpha$ obeys a stable distribution, independent of experimental runs typically of different avalanche sizes. The distribution is broad and nonsymmetric with zero mean. Moreover, after analyzing its spatial distribution at each instant, we find that it is randomly distributed and has no spatial correlation with $v$ by comparing Fig.~\ref{fig:figure2}(c) and Fig.~\ref{fig:figure4}(b). Lastly, from the time evolution of a randomly selected particle's velocity and the angle $\alpha$, no obvious temporal correlation is observed in Fig.~\ref{fig:figure2}(a) and Fig.~\ref{fig:figure4}(a). Therefore, we believe that the two quantities $\alpha$ and $v$ are uncorrelated.

The above expression will predict a strong correlation between $\delta F$ and the summation of $v^2$. We note that if the collision time is a constant then $\langle \delta F\rangle$ is proportional to $\sum_i \langle v_i\rangle$ instead. To verify this, we plot the summation of $v$ (inset) and $v^2$ (main plain) versus time in Fig.~\ref{fig:figure8} (b). After calculation, the correlation between $\delta F$ and $\sum v$ is $0.942$, and between $\delta F$ and $\sum v^2$ is $0.925$, which agrees with the experiment, as displayed in Table~\ref{table:table2}. The above two types of strong correlations indicate the existence of two possible time scales with one inversely proportional to the typical velocity, i.e. $D/\langle v \rangle$,  and the other a constant as shown above. One natural constant time scale can be estimated from $\sqrt{D/g}$, where $D$ is the average diameter of particle and $g$ is the gravitational constant. These two time scales are, however, comparable as estimated in our experiment.
%table 2
\begin{table}%\footnotesize
{
\caption{\label{table:table2}
The statistics of the degree of linear temporal correlations between macroscopic variables in ten different runs. Here `ccf' means the correlation coefficient of the linear fitting.
}
\scriptsize
\begin{tabular}{|c|c|c|c|c|c|c|c|c|c|c|c|c|c|}
\hline
\multicolumn{12}{|c|}  {avalanche number $n_0$} &&\\
\hline
  & $n_0$ & 1 & 2 & 3 & 4 & 5 & 6 & 7 & 8 & 9 & 10 & mean & std \\ \hline
\multirow{3}{*}{ccf}
                & $\delta F-V$ &  0.93  & 0.96  & 0.89 & 0.92 & 0.90 & 0.90 & 0.95 & 0.93 & 0.96 & 0.94 & 0.93 & 0.024	\\ 	
                & $\delta F-V^2$ & 0.91 & 0.92 & 0.91 & 0.90 & 0.92 & 0.91 & 0.94 & 0.90 & 0.93 & 0.94 & 0.92 & 0.015	\\  \hline
\end{tabular}
}
\end{table}

The combination of Fig.~\ref{fig:figure3} and Fig.~\ref{fig:figure8} points to a coherent picture of a plausible mechanism of the particle entrainment. When the kinetic energy of the flowing particles increases, the velocity fluctuation increases; as a consequence, the stress fluctuation in the particle configurations under the surface flowing particles increases. Therefore more particles in the stable particle configuration will be fluidized, a phenomenon of particle entrainment \cite{christen2010ramms,hungr2004entrainment,iverson2011positive}. In the inset of Fig.~\ref{fig:figure8} (a), the dynamics of the total number of particle N indeed confirms the above scenario. Note that in more realistic geological processes, the saturated water content in the stable particle configuration can make significant contributions to the particle entrainment \cite{iverson2011positive}. In addition, our system consists of circular shaped particles without cohesive forces. It is known from the literature that the change of the particle roughness or the inclusion of the cohesion force may change the dynamics of the avalanches \cite{borzsonyi2005two, tegzes2002avalanche}. It remains an open question on how the main results in Fig.~\ref{fig:figure3} and Fig.~\ref{fig:figure8} would be changed if more complex particle shapes and interactions are considered. In particular, we believe the statistics of $\alpha$ is only a function of the particle properties.

\section{Discussion}

To conclude, we have analyzed the nonaffine particle motion in the granular avalanches and we find that there are strong correlations between velocity fluctuations and velocity during granular avalanche: at the particle scale these two quantities correlate in both the spatial and the temporal domains; at the macroscopic scales, the nonaffine motion increases as the total kinetic energy increases. By simulating a steady shear flow with an independent distribution of velocity orientation angle $\alpha$ with respect to the inclination, we can qualitatively reproduce some main results of experimental measurements. These results may not be a direct consequence of the STZ theory though T1 events have been successfully observed during avalanche for the first time in a 2D granular experiment. In addition, by adapting an analytical model from impact dynamics, we find that the independent distribution of $\alpha$ will lead to a strong correlation between the total velocity magnitude or kinetic energy of the system and the stress fluctuations of the system. Our findings will be important for the further development of theoretical descriptions of the dynamics of the granular avalanche. Our findings also provides a coherent physical picture of a plausible mechanism of particle entrainment in a simple granular system.

\section{Methods}

\begin{figure}
\centerline{\includegraphics[width=0.5\textwidth]{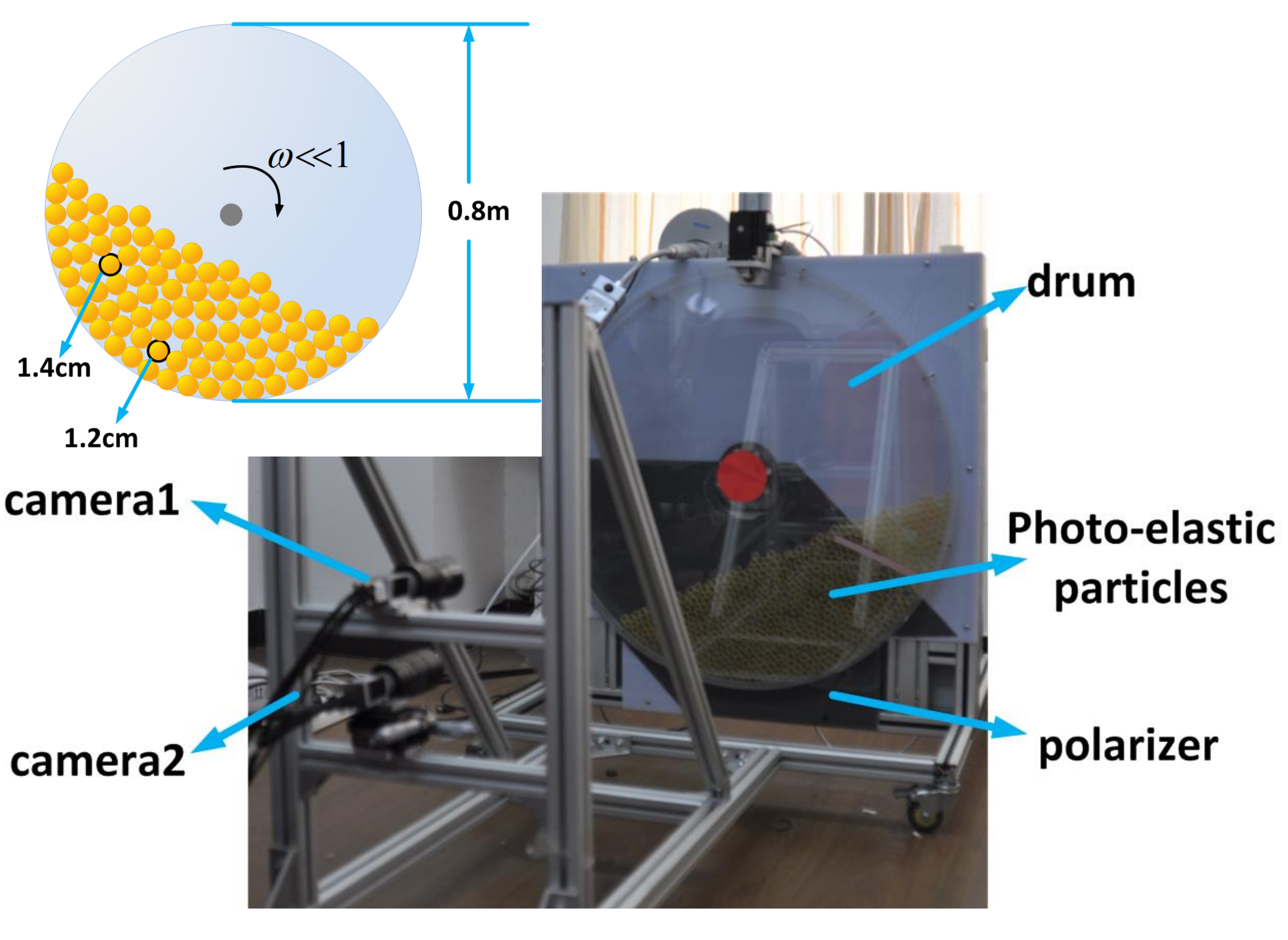}}
\caption{\label{fig:figure1} A snapshot (main panel) and a schematic (inset) of the experimental setup. }
\end{figure}

The experimental setup mainly consists of a thin rotating drum of a diameter of $80cm$ and a width of $1cm$, as shown in Fig.~\ref{fig:figure1}. The drum is driven by a stepping motor to rotate at a slow speed of $1/15$ revolutions per minute. The construction of the drum is composed of two Plexiglas plates with a gap of $8mm$ and with the inner surfaces coated to eliminate the accumulation of electrostatic charges. A thin layer of $736$ photo-elastic disks is sandwiched in between the two plates. These quarter-inch-thick disks are bi-disperse with a large size of $1.4cm$ in diameter and a small size of $1.2cm$ in diameter and with a size ratio of $1:1$. They are machined from PSM4 materials from Vishay and only fill less than half of the space of the drum. In front of the drum, two high-speed cameras (2048$\times$2048 pixels and with a frame rate up to 180/sec) are mounted to continuously record images during the avalanche. In front of the lens of one camera, a piece of polarizer is placed in order to capture the force-chain network simultaneously.

To improve the statistics, the experiment has been repeated for ten times following the same protocol. Results of different runs are similar. In this paper, we will present experimental results in two groups: the first group includes results from a randomly selected experimental run in order to show details of the avalanche process; the second group consists results of all ten runs in order to present the statistics of the granular avalanche.

%\bibliography{dynamical}

{\bf Acknowledgments}\\
J.Z. thanks the helpful discussion with Y. Wang and N. Zheng. J.Z. thanks J. Dijksman for the critical reading of the manuscript. J.Z. acknowledges support from the award of the Chinese 1000-Plan (C) fellowship, Shanghai Pujiang Program (13PJ1405300), and National Natural Science Foundation of China (11474196).\\
{\bf Author Contributions}\\
J.Z. and Z.W. wrote and reviewed the main manuscript text, and Z.W. prepared the figures 1-8.\\
{\bf Additional Information}\\
Competing financial interests: The authors declare no competing financial interests.

\end{document}